 \documentclass[prd,twocolumn,showpacs,preprintnumbers,amsmath,amssymb]{revtex4}
\usepackage{graphicx}
\usepackage{epsfig}
\usepackage{rotating}
\usepackage{dcolumn}
\usepackage{bm}
\begin{document}
\title{AXIAL VECTOR CURRENT MATRIX ELEMENTS AND QCD SUM RULES}
\author{J. Pasupathy}
\email{jpcts@cts.iisc.ernet.in}
\affiliation{Center for Theoretical Studies, IISc, Bangalore, India} 
\author{Ritesh K. Singh}
\email{ritesh@cts.iisc.ernet.in}
\affiliation{Center for Theoretical Studies, IISc, Bangalore, India} 
\begin{abstract}
The matrix element of the isoscalar axial vector current,
$\bar{u}\gamma_\mu\gamma_5u + \bar{d}\gamma_\mu\gamma_5d $, between nucleon 
states is computed using the external field QCD sum rule method. The 
external field induced correlator, $\langle0|\bar{q}\gamma_\mu\gamma_5q|0
\rangle$,  is calculated from the spectrum of the isoscalar axial vector meson 
states. Since it is difficult to ascertain, from QCD sum rule for hyperons, 
the accuracy of validity 
of flavour SU(3) symmetry in hyperon decays when strange quark mass is taken 
into account, we rely on the empirical validity of Cabbibo theory to dertermine
the matrix element 
$\bar{u}\gamma_{\mu}\gamma_5 u + \bar{d}\gamma_{\mu}\gamma_5 d - 2 
\bar{s}\gamma_{\mu}\gamma_5 s$ between nucleon states. Combining with our 
calculation  of $\bar{u}\gamma_{\mu}\gamma_5 u + \bar{d}\gamma_{\mu}\gamma_5 d$ 
and the well known nucleon $\beta$-decay constant allows us to determine 
$\left\langle p,s\right| \frac{4}{9}\bar{u}\gamma_{\mu}\gamma_5 u +
\frac{1}{9}\bar{d}\gamma_{\mu}\gamma_5 d +
\frac{1}{9}\bar{s}\gamma_{\mu}\gamma_5 s \left|p, s\right\rangle$
occuring in the Bjorken sum rule. The result is in reasonable agreement with
experiment. We also discuss the role of the anomaly in maintaining flavour
symmetry and validity of OZI rule.
\end{abstract}
\maketitle
\noindent
\section{Introduction}
\label{one}
The determination of flavour singlet and non-singlet axial vector matrix
elements between nucleon states is of considerable interest theoretically and 
experimentally. While non-singlet currents appear in nucleon and hyperon beta
decays, the linear combination
\begin{equation}
\left\langle p,s\right| \frac{4}{9}\bar{u}\gamma_{\mu}\gamma_5 u + 
\frac{1}{9}\bar{d}\gamma_{\mu}\gamma_5 d +
\frac{1}{9}\bar{s}\gamma_{\mu}\gamma_5 s \left|p, s\right\rangle = 
-s_{\mu} G_{Bj}
\label{bj}
\end{equation}
appears in the integral of the first moment of the polarised structure function
$g_1(x)$ in the sumrule of Bjorken \cite{bj}. Here $|p, s\rangle$ denotes 
proton state of 
momentum $p$ and polarisation $s_\mu$. We have introduced the constant $G_{
Bj}$ to denote the linear combination of axial vector current with coefficients
equal to square of quark charges. Nearly three decades ago Ellis and Jaffe
\cite{ej} using a simple minded application of the OZI rule, set the strange 
quark current matrix element between proton states to be zero, which 
immediately enabled them to write 
\begin{equation}
G_{\rm Bj} = \frac{1}{6} G_A + \frac{5}{18} G_8
\label{gbj}
\end{equation}
where, $G_A$ is the isovector axial vector coupling occuring in nucleon beta
decay and $G_8$ is the octet current coupling. Eqn.(\ref{gbj}) is in 
disagreement with experiment \cite{adams} if SU(3) flavor symmetry is a good 
symmetry \cite{leader}. Already in 1979 Gross, Trieman and Wilczek 
\cite{gross} had pointed out that because the light quark masses  are unequal,
i.e.
$$ \frac{m_d-m_u}{m_u+m_d} = O(1),$$
one should expect large violations of Isospin symmetry in the Bjorken sumrule
if anomaly is neglected. The matrix elements of the anomaly between vacuum
and Goldstone states $|\pi\rangle$ and $|\eta\rangle$  are not zero and 
play a crucial role in maintaining flavour symmetry. As we shall  discuss 
further in Section \ref{two}, this at
once implies that Ellis-Jaffe assumptions of simulataneous validity of
SU(3) flavour symmetry and OZI rule are mutually incompatible. The octet
current has no anomaly while the  singlet  does.

The computation of matrix elements of the axial current in Eqn.(\ref{bj}) 
is clearly a problem in QCD. This can be addressed using the  external field 
method introduced in the context of QCD sumrules by Ioffe and Smilga
\cite{ioffe} and Balitsky and Yung \cite{iib} in 1983 to calculate the magnetic 
moments and which has since then been used for numerous other matrix elements as
well. In fact computation of the axial vector current matrix elements has been
done using these methods in ref.\cite{pisma}-\cite{ago}.

The reasons for reconsidering the earlier QCD sum rule determination
of $G_{Bj}$ are as follows. 
\begin{enumerate}
\item Besides the usual Lorentz invariant 
chiral and gluon condensates present in QCD vacuum, additional
non-invariant correlators $\langle0|\bar{q}\gamma_\mu\gamma_5 q|0\rangle$   
induced by the external field, enter
the sumrules. If in Eqn.(1), on the left hand side, we had only    
flavour non-singlet currents, the corresponding external field induced 
correlator of dimension three,
\begin{eqnarray}
&&\langle0|\bar{u}\gamma_\mu\gamma_5 u - \bar{d}\gamma_\mu\gamma_5
d|0\rangle |_{\rm Ext. \ Field} \hspace{1.0cm}{\rm and}\\
&&\langle0|\bar{u}\gamma_\mu\gamma_5 u + \bar{d}\gamma_\mu\gamma_5 d
- 2 \bar{s}\gamma_\mu\gamma_5 s |0\rangle|_{\rm Ext. \ Field}
\end{eqnarray}
for example, can be found using the ward identity and PCAC.

The determination of either the isoscalar current or SU(3) singlet current
correlators 
\begin{eqnarray}
&&\langle0|\bar{u}\gamma_\mu\gamma_5 u + \bar{d}\gamma_\mu\gamma_5
d|0\rangle |_{\rm Ext. \ Field}  \\
&&\langle0|\bar{u}\gamma_\mu\gamma_5 u + \bar{d}\gamma_\mu\gamma_5 d
+ \bar{s}\gamma_\mu\gamma_5 s |0\rangle|_{\rm Ext. \ Field}
\end{eqnarray}
with the help of Ward identities are no longer simple since they involve the
gluon anomaly. In earlier works some authors \cite{bely,gupta,henley} used 
simply the values 
of the non-singlet current induced correlators, as an expedient measure, while
Ioffe and Khodjamirian \cite{yu} found inconsistencies with SU(3) flavor
symmetry in their calculations of the correlator in
Eqn.(6). Indeed later in ref.\cite{ago}, the correlator Eqn.(6) was treated as 
an unknown free parameter and its value fixed by using the experimental
value of $G_{Bj}$.  In the present work we follow a different procedure.
The external field induced correlator corresponding to the isoscalar
current Eqn.(5) is determined directly from the axial vector meson spectrum and
we verify that similar determination of the nonsinglet isovector
current induced correlator Eqn.(3) indeed yields a value consistent with
Ward identity and PCAC. 
\item We also clarify the differences in the calculations of the Wilson 
coefficients between ref.\cite{chiu} and \cite{gupta} on one hand and ref. 
\cite{bely} on the other.
\item In the analysis of sumrules,
we use a value of the QCD scale parameter $\Lambda$  consistent with present
data, while earlier works used a significantly lower value.
\end{enumerate}

This paper is organised as follows. In the next section we briefly recapitulate 
the arguments of Gross, Treiman and Wilczek, regarding the role of the anomaly
in maintaining the flavour symmetry. Although the anomaly  is superficially a
flavor singlet its matrix elements between the vacuum and the Goldstone state
$|\pi^0\rangle$ and $|\eta\rangle$, are not zero. We also make a brief 
digression on the OZI rule. In section \ref{three} we give a summary of the 
external field method, and an appendix explains the differences between ref.
\cite{chiu} and \cite{bely} in the computation of the Wilson coefficients. 
In Section \ref{four} we outline the determination of the external field
induced vacuum correlators which is used in Section \ref{five} to determine 
the isoscalar matrix element and we end with a brief discussion.
\section{Flavour Symmetry, OZI, and The Anomaly}
\label{two}
We briefly recall the argument of ref \cite{gross}. If we ignore the anomaly 
then we have, for the divergence of isoscalar current
\begin{eqnarray}
\partial^{\mu}\left[ \bar{u}\gamma_{\mu}\gamma_5 u +
\bar{d}\gamma_{\mu}\gamma_5 d \right]&=&i(m_u+m_d)\left[
\bar{u}\gamma_5 u + \bar{d}\gamma_5 d \right] \nonumber\\ 
&+&i(m_u-m_d)\left[\bar{u}\gamma_5 u - \bar{d}\gamma_5 d \right]\nonumber\\
\end{eqnarray}
We also have, for the isovector 
\begin{eqnarray}
F_{\pi}m_{\pi}^2 \phi_{\pi0}&=&\partial^{\mu}\left[ \bar{u}\gamma_{\mu} 
\gamma_5 u - \bar{d}\gamma_{\mu}\gamma_5 d \right]\nonumber\\
&=&i(m_u+m_d)\left[ \bar{u}\gamma_5 u - \bar{d}\gamma_5 d\right] \nonumber\\
&+&i(m_u-m_d)\left[ \bar{u}\gamma_5 u + \bar{d}\gamma_5 d \right]
\end{eqnarray} 
Combining these using PCAC one gets
\begin{eqnarray}
\langle N|\partial^{\mu}( \bar{u}\gamma_{\mu} \gamma_5 u + \bar{d}\gamma_{\mu}
 \gamma_5 d)|N\rangle_{I=1} = \nonumber \\
\frac{m_u-m_d}{m_u+m_d}\langle N| \partial^{\mu}
(\bar{u}\gamma_{\mu} \gamma_5 u - \bar{d}\gamma_{\mu}\gamma_5 d)|N\rangle_{I=1}
\label{pcac}
\end{eqnarray}
where the subscript $I=1$ on the nucleon states denotes the difference between
proton and neutron matrix elements
$$\langle N|{\cal O}|N\rangle_{I=1}= \langle p|{\cal O}|p\rangle - \langle n|
{\cal O}|n\rangle.$$
Eqn.(\ref{pcac}) implies a large violation of isospin in Bjorken sum-rule since
$$\frac{m_d-m_u}{m_u+m_d} = O(1).$$
This conclusion is avoided by noting that one has ignored the anomaly.
In Eqn.(7) one should write 
\begin{eqnarray}
\partial^{\mu}\left[ \bar{u}\gamma_{\mu}\gamma_5 u +
\bar{d}\gamma_{\mu}\gamma_5 d \right]&=&i(m_u+m_d)\left[
\bar{u}\gamma_5 u + \bar{d}\gamma_5 d \right] \nonumber\\
&+&i(m_u-m_d)\left[\bar{u}\gamma_5 u - \bar{d}\gamma_5 d \right]\nonumber\\
&+&2\frac{g^2}{16\pi^2}G_{\mu\nu}^a\tilde{G}_{\mu\nu}^a
\end{eqnarray}                                                          
where $\tilde{G}_{\mu\nu}^a = \frac{1}{2}\epsilon_{\mu\nu\alpha\beta}
G^{a\alpha\beta}$. Using a Sutherland type argument  it is derived in ref \cite{gross} 
\begin{eqnarray}
i\langle 0|(m_u\bar{u}\gamma_5 u + m_d \bar{d}\gamma_5 d)|\pi^0\rangle + 
\langle 0|\frac{g^2}{16\pi^2} G_{\mu\nu}^a\tilde{G}_{\mu\nu}^a|\pi^0\rangle =
0\nonumber\\
\end{eqnarray}
and 
\begin{eqnarray}
\langle 0|2 i(m_q\bar{q}\gamma_5 q )|\pi^0\rangle + 
\langle 0|\frac{g^2}{16\pi^2} G_{\mu\nu}^a\tilde{G}_{\mu\nu}^a|\pi^0\rangle = 0
\end{eqnarray}
where, $q = s, \ c,...$ etc. Using again a PCAC argument they\cite{gross} obtain
\begin{eqnarray}
2 i\langle 0| m_u \ \bar{u}\gamma_5 u + m_d \ \bar{d}\gamma_5 d 
|\pi^0\rangle &=& \frac{m_u-m_d}{m_u+m_d} \ F_{\pi} \ m_{\pi}^2\sqrt{2} 
\nonumber\\ 
&=&-2 \langle 0|\frac{g^2}{16\pi^2}G_{\mu\nu}^a\tilde{G}_{\mu\nu}^a|\pi^0\rangle
\nonumber\\
&=& 4i \langle 0|m_s \ \bar{s}\gamma_5 s |\pi^0\rangle \nonumber\\
&=& 4i \langle 0|m_c \ \bar{c}\gamma_5 c |\pi^0\rangle \nonumber\\
&=& 4i \langle 0|m_b \ \bar{b}\gamma_5 b |\pi^0\rangle \ ... 
\label{ud}
\end{eqnarray}
In other words the matrix elements of the anomaly are far from being
a flavour singlet. It was pointed out by Novikov et.al \cite{van} that the 
matrix elements of the anomaly between vacuum and $|\eta\rangle$ is again not 
zero.

Writing 
\begin{equation}
\langle0|\bar{u}\gamma_{\mu}\gamma_5 u + \bar{d}\gamma_{\mu}\gamma_5 d 
-2 \bar{s}\gamma_{\mu}\gamma_5 s |\eta\rangle = i\sqrt{6} F_\pi  k_\mu,
\end{equation}
and taking its diveregence one gets
\begin{equation}
\langle0|-4i m_s\bar{s}\gamma_5 s |\eta\rangle = \sqrt{6}F_\pi m_\eta^2,
\label{ss}
\end{equation}
where we have ignored $m_u$ and $m_d$ in comparison with $m_s$. For the singlet 
current one may write 
\begin{equation}
\langle0|\bar{u}\gamma_\mu\gamma_5 u + \bar{d}\gamma_\mu\gamma_5 d +
\bar{s}\gamma_\mu\gamma_5 s |\eta\rangle = i f_1 k_\mu
\end{equation}
If SU(3) flavour symmetry were exact, as happens when all light quark
masses are neglected, $f_1 = 0$. On the other hand $F_\pi$ remains finite 
in the chiral limit of massless quarks. We can expect then $F_\pi >> f_1$.
Setting $f_1 = 0$, it is then easy to obtain from Eqns. (15) and (16) 
cf.\cite{van},
\begin{equation}
\langle0| \ \frac{3\alpha_s}{4\pi} \ G_{\mu\nu}^a\tilde{G}_{\mu\nu}^a
|\eta\rangle = \sqrt{\frac{3}{2}}F_\pi m_\eta^2
\label{gg}
\end{equation}
We learn from Eqns.(\ref{ud}) and (\ref{gg}) above that the matrix elements of 
the anomaly between the vacuum and non-singlet goldston boson $|\pi^0\rangle$ 
and $|\eta\rangle$ is not zero but proportional to quark mass differences.


We note that Ioffe and Shifman \cite{shif} using
Eqns. (\ref{ud}) and (\ref{gg}) above obtained the result
\begin{eqnarray}
r&=&\frac{\Gamma(\psi(2s)\to J/\psi(1s)\pi^0)}{\Gamma(\psi(2s)\to J/\psi(1s) 
\eta)} \nonumber \\
&=&3\left(\frac{m_d-m_u}{m_d+m_u}\right)^2\left(\frac{m_\pi}{m_\eta}
\right)^4 \left(\frac{p_\pi}{p_\eta}\right)^3
\end{eqnarray}
Experimentally one has $r=(3.07 \pm 0.70)\times 10^{-2}$. We use this to
find $m_u/m_d$ and obtain
$$ \frac{m_u}{m_d} = 0.44 \pm 0.07$$
which is consistent with values obtained using entirely different inputs; Gao
\cite{gao} $m_u/m_d = 0.44$ and Leutwyler \cite{leut}  $m_u/m_d = 
0.553\pm0.043$.

Encouraged by the agreement between Eqn.(\ref{gg}) and experiments let us
consider the matrix element
\begin{equation}
\langle0|\bar{s}\gamma_\mu\gamma_5 s|\eta\rangle = i f_s k_\mu
\end{equation}
Taking the divergence 
\begin{eqnarray}
\langle0|\partial^\mu\bar{s}\gamma_\mu\gamma_5 s|\eta\rangle&=&\langle0|
2im\bar{s}\gamma_5 s + \frac{\alpha_s}{4\pi}G_{\mu\nu}^a \tilde{G}_{\mu\nu}^a
|\eta\rangle \nonumber\\
&=& f_s m_\eta^2
\label{ss1}
\end{eqnarray}
Using Eqns.(\ref{ss}) and (\ref{gg}) we have
\begin{equation}
f_s = \frac{-\sqrt{6}}{3} F_\pi
\label{fsfp}
\end{equation}
Now it is known that the Goldberger-Treiman relation
\begin{equation}
\sqrt{2} m_N G_A = g_{\pi N} F_\pi
\label{gt}
\end{equation}
is accurate to a few percent and is exact in the chiral limit of massless
quarks. Corrections to it have the structure \cite{hrp}
\begin{eqnarray}
1-\frac{\sqrt{2} m_N G_A}{g_{\pi N} F_\pi} = \Delta = C_1 m_\pi^2 + C_2 m_\pi^4
ln(m_\pi^2) + ..
\label{dgt}
\end{eqnarray}
In other words, GT relation is obtained by retaining only the Goldstone pole and
discarding the continuum contribution in the dispersion integrands. In the
chiral limit, equations analogous to Eqns.(\ref{gt}) and (23) also hold good 
for the nucleon matrix element of the octet current with $G_A$ replaced by
$G_U+G_D-2G_S$ and $g_{\pi N}$ replaced by $g_{\eta N}$ etc. with leading
corrections proportional to quark mass as in Eqn(\ref{dgt}). If we now naively 
extend these dispersion relation considerations to the 
nucleon matrix element of the strange quark current $\bar{s}\gamma_\mu \gamma_5
s$, retaining only the $\eta$-pole and discard the continuum as well as $\eta'$
pole we immediately obtain from Eqn.(14), (19) and (\ref{fsfp})
\begin{equation}
G_S = \frac{-1}{3} (G_U+G_D-2G_S)
\end{equation}
or
\begin{equation}
G_U+G_D+G_S = 0
\end{equation}
which is same as the Skyrme model result \cite{sb}.

Our main point in this section is that OZI cannot be applied naively.
The anomaly is important to avoid gross violations of SU(3) flavour symmetry 
and matrix elements of the anomaly  are not flavour symmetric.

It is worth emphasising that OZI rule violates unitarity, a cardinal
property of all S-matrix elements. Corrections  to OZI rule can be estimated 
using unitarity and they are very much process dependent \cite{jp}. Charmonium 
decays illustrate the point. For example,
\begin{eqnarray}
{\rm B.R.}(J/\psi(1s) \to \rho\pi) &=& 1.27 \times 10^{-2},\hspace{1cm} {\rm 
while}\nonumber\\
{\rm B.R.}(\psi(2s) \to \rho\pi) &\le& 8.3 \times 10^{-5} \nonumber
\end{eqnarray}
despite the fact $\Gamma_{tot}(\psi(2s)) = 277$ keV is just a factor of three 
larger than $\Gamma_{tot}(J/\psi(1s)) = 87$ keV. Again in the decay of 
$J/\psi(1s)$
into light mesons, SU(3) flavor symmetry works better in pseudoscalar vector
decays than in vector tensor decays. Also the decay mode  $J/\psi(1s)\to 
\phi\pi\pi$ is not doubly suppressed as one might naively expect. These 
emphasise the point that
corrections to OZI rule are to be studied individually for each matrix element
\cite{jp} and there is no universal principle which tells a priori how good 
is the OZI rule for any specific matrix element.
%
%
%
%
\section{QCD sumrule with an external field}
\label{three}
We shall follow closely the notations of Ioffe \cite{bli}. We consider the 
nucleon correlator in an external field
\begin{equation}
\Pi(p,A_\mu) = \left.i\intop d^4x \ e^{i p.x}\langle0|T\{\eta(x),\bar{\eta}(0)
\}|0\rangle\right|_{A_\mu}
\label{PiA}
\end{equation}
where $\eta(x)$ is the nucleon current
\begin{equation}
\eta(x) = \epsilon^{abc}u^a(x) \ C\gamma_\mu u^b(x)\gamma^\mu\gamma_5 d^c(x)
\label{eta}
\end{equation}
with proton quantum numbers; $u^a, d^b$ are quark fields and $a,b,c$ are color
indices. $A_\mu$ refers to constant external field. To compute the matrix 
element of the current $j_\mu^5$ between a proton state $\langle p|j_\mu^5 |p\rangle$ 
one adds a term 
\begin{equation}
\Delta{\cal L} = j^5_\mu A^\mu
\end{equation}
to the Lagrangian and evaluates $\Pi(p,A_\mu)$ in Eqn.(\ref{PiA}) upto terms 
linear in $A_\mu$. In ref \cite{pisma}-\cite{yu} the following four different 
currents were considered.
\begin{eqnarray}
j^5_\mu &=& \bar{u}\gamma_\mu\gamma_5u - \bar{d}\gamma_\mu\gamma_5d\hspace{2cm}
\text{isovector}\\
j^5_\mu &=& \bar{u}\gamma_\mu\gamma_5u + \bar{d}\gamma_\mu\gamma_5d\hspace{2cm}
\text{isoscalar}\\
j^5_\mu &=& \bar{u}\gamma_\mu\gamma_5u + \bar{d}\gamma_\mu\gamma_5d 
-2\bar{s}\gamma_\mu\gamma_5s \hspace{1cm} \text{octet}\\
j^5_\mu &=& \bar{u}\gamma_\mu\gamma_5u + \bar{d}\gamma_\mu\gamma_5d 
+\bar{s}\gamma_\mu\gamma_5s \hspace{0.3cm} \text{SU(3) singlet}
\end{eqnarray}
Bearing in mind that the corrections to OZI rule are a priori unknown
and can be large, we shall consider only the isoscalar current Eqn.(30) here,
since the nucleon current of Ioffe in Eqns.(26-27) above has
only up and down quark fields, and therefore couples to the $\bar{s}\gamma_\mu
\gamma_5 s$ term in the octet current, Eqn.(31), or singlet, Eqn.(32),only 
through gluonic corrections that is by corrections to OZI rule.

In deriving the QCD sumrules one must
take into account, external field induced correlators . We define
\begin{eqnarray}
\langle0|\bar{q}\gamma_\mu\gamma_5 q|0\rangle|_{A_\mu} &=& F A_\mu\\
\langle0|g_s\bar{q}\frac{\lambda^a}{2}\tilde{G}^a_{\rho\mu}\gamma^\rho 
q|0\rangle|_{A_\mu} &=& H A_\mu
\end{eqnarray}
It is clear that the constants $F$ and $H$
are in general different corresponding to the different $\Delta{\cal L}$
introduced in Eq.(28-32). The constant $F$ can be obtained from
\begin{eqnarray}
\langle0|\bar{q}\gamma_\mu\gamma_5 q|0\rangle|_{A_\mu} = i\int 
d^4x \langle0|T (\Delta{\cal L},\bar{q}\gamma_\mu\gamma_5 q)|0\rangle
\end{eqnarray}
and the constant $H$ from
\begin{eqnarray}
\langle0|g_s\bar{q}\frac{\lambda^a}{2}\tilde{G}^a_{\rho\mu}\gamma^\rho q|0
\rangle |_{A_\mu}\nonumber\\
 = i\int d^4x \langle0|T (\Delta{\cal L},
g_s\bar{q}\frac{\lambda^a}{2}\tilde{G}^a_{\rho\mu}\gamma^\rho q)|0\rangle
\end{eqnarray}
Complete details of the calculation of $\Pi(p,A_\mu)$ in Eqn.(\ref{PiA})
can be found in ref.\cite{pisma,chiu,bely,gupta,henley,yu}
for both the isovector and isoscalar currents. For the isoscalar axial
vector matrix element the sum rule then reads
\begin{eqnarray}
&&\frac{-M^6}{L^{4/9}}E_2+\frac{16\pi^2}{3} \ F \ \frac{M^4}{L^{4/9}}E_1 
\nonumber\\
&&+\frac{b}{4}\frac{M^2}{L^{4/9}}E_0 +\frac{16\pi^2}{3} \ H \ 
\frac{M^2}{L^{8/9}} E_0 - \frac{4}{9} a^2 L^{4/9} \nonumber\\
&&= \tilde{\lambda}_N^2 (G+AM^2) exp[-m_N^2/M^2]
\label{rule}
\end{eqnarray}
Here $m_N$ is the nucleon mass and $M^2$ is the Borel mass variable. In the 
right hand side, $\tilde{\lambda}_N$ is defined by
$$\langle0|\eta(x)|p\rangle = \lambda_N v(p) \hspace{1cm}{\rm and}\hspace{1cm}
\tilde{\lambda}_N^2 = \frac{\lambda_N^2}{32\pi^2}$$
$G_U + G_D = G$  is the isoscalar axial vector current matrix element. The term
$A M^2$ arises from the fact that in the presence of external
field there are non-diagonal transition between nucleon and excited
states. In the left hand side
\begin{eqnarray*}
L=\frac{ln(M/\Lambda)}{ln(\mu/\Lambda)}, \ a =
-(2\pi)^2\langle\bar{q}q\rangle, \ b = \langle g_s^2G^a_{\mu\nu}G^a_{\mu\nu}
\rangle,\\
E_0 = 1-e^{-W^2/M^2}, \\ E_1 = 1 - (1+W^2/M^2) \ e^{-W^2/M^2}, \\
E_2 = 1-(1+W^2/M^2+W^4/2M^4) \ e^{-W^2/M^2}
\end{eqnarray*}
$\mu$ is the renormalization scale, which we take to be 1 GeV, and $\Lambda$ is 
the QCD scale which for 3 flavor case is 247 MeV \footnote{This corresponds to 
$\alpha_s (1 \ GeV) = 0.5$ with one loop beta function with three flavors.}.

Although the sum rule has been written down earlier by several authors
the purpose of reconsidering it here are the following.
\begin{enumerate}
\item It is clear the constants $F$ and $H$ should be determined from 
other sumrules or different
considerations before we can use it to find $G$ in Eqn.(37). As will be 
explained in detail in the next section, for the isovector case, it is 
relatively
easy to determine them using PCAC. For the singlet currents  one must
take into account the anomaly. A decade ago Ioffe and Khodjamirian \cite{yu}
attempted to compute  $F$ using the Ward identity and sum rules for the 
divergence of the SU(3) singlet current. They found inconsistency with 
SU(3) flavour symmetry chiefly because of the large difference between the 
strange quark mass and the up or down quark mass. In later works 
\cite{bli,ago} this lead Ioffe and his collaborators to use the sumrule,
Eqn.(37), to find $F$ using the experimental value of $G_{Bj}$ in
Eqn.(1). In contrast here for the
isoscalar matrix element we shall determine the constant  $F$ from the
spectrum of isoscalar axial vector mesons, and use that value to determine
from Eqn.(37) the value of $G$ and therefore $G_{Bj}$.
\item  The external field Lagrangian, modifies the propagation of the current 
$\eta(x)$, in the QCD vacuum in two ways. One in which the external field 
directly couples to the fields in $\eta(x)$. The other, in which the external
field modifies the vacuum as represented by the induced correlators  $F$ and 
$H$, which appear in the second and fourth terms in the sum rule, Eqn.(37). 
Now consider the difference 
between the three $\Delta{\cal L}$ in Eqns.(30), (31), and (32).
As long as gluon loops are not included, the additional term $\bar{s} 
\gamma_\mu\gamma_5sA^\mu$ will not couple to the $u$ and $d$ fields in 
$\eta(x)$. The strange quark term will only affect the constants $F$ and $H$ 
in Eqn.(35) and (36). 
Putting it differently with this assumption, the same sum rule, Eqn.(37), is 
valid for all the three currents, Eqn.(30)-(32) namely, the isoscalar, octet 
and SU(3) singlet with only the external field induced correlators $F$ and $H$
being different. Correspondingly if we were to compute only the matrix
element of $\langle p,s|\bar{s}\gamma_\mu\gamma_5s|p,s\rangle$ then we would 
consider, in place of Eqn.(28), the external field Lagrangian
$$\Delta{\cal L} = \bar{s}\gamma_\mu\gamma_5sA^\mu$$ 
and the corresponding sum rule for the strange quark matrix element in Eqn.(37)
will have, in its left hand side, only the second and fourth terms due to
modifications of the vacuum and all other terms will be set to zero. In the
light of our discussion of OZI rule, one may then expect that the sum rule 
Eqn.(37) will work better for the isoscalar matrix element than for the octet
or SU(3) singlet.
\item
The sum rule, Eqn.(37), differs from those of ref.\cite{bely,ago} also in the 
coefficient of the third and fourth terms due to difference in the calculation 
of Wilson coefficients. This point is elaborated in the Appendix.
\end{enumerate}
\section{External Field induced correlators}
\label{four}
We now turn to the computation of the constants $F$ and $H$  defined
in Eqns.(33-36). First consider the isovector case with the current defined in 
Eqn.(29) and the corresponding correlator
\begin{eqnarray}
\Pi_{\mu\nu}^{I=1}&=&\frac{i}{2}\intop d^4x e^{i q.x} \nonumber\\&& \langle0|
T(\bar{u}(x)\gamma_\mu\gamma_5 u(x) - \bar{d}(x)\gamma_\mu\gamma_5 d(x),
\nonumber\\
&&\bar{u}(0)\gamma_\nu\gamma_5 u(0) - \bar{d}(0)\gamma_\nu\gamma_5 d(0) )|0\rangle
\end{eqnarray}
One can write
$$\Pi_{\mu\nu}^{I=1} = -\Pi^{I=1}_1(q^2) g_{\mu\nu}+\Pi^{I=1}_2(q^2)
q_\mu q_\nu$$
where the coefficient of $g_{\mu\nu}$, $\Pi_1$, gets contribution only from 
spin $1^+$ states while the coefficient of $q_\mu q_\nu$, $\Pi_2$, has 
contributions form both $1^+$  and $0^-$ states.

It is easy to see, using the gauge condition, $A_\mu q^\mu = 0$ and Eqns.(28), 
(32) and (34) that 
\begin{equation}
F(I=1) = - \Pi^{I=1}_1(q^2=0)
\end{equation}
This can be easily  evaluated using the ward-identity which reads
\begin{eqnarray}
&-&\Pi_1^{I=1}(q^2) \ q^2 + \Pi_2^{I=1}(q^2) \ q^4 \nonumber \\
&=& (m_u+m_d) \langle0|\bar{u} u + \bar{d} d|0\rangle \nonumber \\
&&-i(m_u+m_d)^2\intop d^4x e^{i q.x} \langle0|T (
\bar{d}\gamma_5 u(x), \bar{u}\gamma_5 d(0) )|0\rangle \nonumber \\
\end{eqnarray}
Isolating the pion pole in $\Pi_2(q^2)$ and pseudoscalar correlation matrix
element in the r.h.s. we can write near the pion pole
\begin{equation}
-\Pi_1^{I=1} (q^2) q^2 + \frac{F_\pi^2q^4}{(m_\pi^2-q^2)} \approx -
F_\pi^2m_\pi^2 + \frac{F_\pi^2 m_\pi^4}{(m_\pi^2-q^2)}
\end{equation}
From which we get
\begin{equation}
\Pi_1^{I=1}(0) = -F_\pi^2
\label{pi0}
\end{equation}
For the isoscalar case we need to consider the analogue Eqn(38)
\begin{eqnarray}
\Pi_{\mu\nu}^{I=0}&=&\frac{i}{2}\intop d^4x e^{i q.x} \nonumber\\ &&\langle0|
T(\bar{u}(x)\gamma_\mu\gamma_5 u(x) + \bar{d}(x)\gamma_\mu\gamma_5 d(x),
\nonumber\\
&&\bar{u}(0)\gamma_\nu\gamma_5 u(0) + \bar{d}(0)\gamma_\nu\gamma_5 d(0) )|0\rangle
\end{eqnarray} 
with
\begin{equation}
F(I=0) = - \Pi^{I=0}_1(q^2=0)
\end{equation}  
However, unlike the isovector case the ward identities are no longer simple 
since
it involves anomaly Eqn.(10) and Eqn.(40) is replaced, in the right hand side,
by considerably more complex set of terms including the anomaly. We shall
therefore not use the ward identity to find $\Pi_1^{I=0}(0)$.

On the other hand $\Pi_1^{I=0}(q^2)$ satisfies a dispersion relation. Using the
operator product expansion for $\Pi_1^{I=0}(q^2)$ the value of $\Pi_1^{I=0}(0)$
can be found by the QCD sum rule method. Analogous procedure of course can be
used for the isovector case as well. This procedure is well known and complete 
details can be found in ref.\cite{see}. Here we shall simply write the final
result. Denoting by $\hat{L}_{M^2}$ the Borel transform \footnote{ The Borel
Transformation of a function $f(q^2)$ is defined by  $\hat{L}_{M^2} f(q^2) =
\frac{(-1)^N(Q^2)^N}{(N-1)!}\left(\frac{\partial}{\partial Q^2}\right)^N
f(q^2) $, $\frac{Q^2}{N} = M^2$ kept fixed.} we have from
\begin{eqnarray}
\Pi_1(q^2) &=& \frac{1}{\pi}\intop\frac{\Im\Pi_1(s) \ ds}{s-q^2} + {\rm
subtractions}\\ \label{pi}
\hat{L}_{M^2}\Pi_1(q^2) &=& \frac{1}{\pi M^2}\intop \Im\Pi_1(s) \ e^{-s/M^2} \
ds 
\label{lpi}
\end{eqnarray}
where $\Im\Pi_1$ denotes imaginary part of $\Pi_1$. Eqn.(\ref{lpi}) is usually 
used to compute the mass of the $f_1$ state when isoscalar correlator is
considered and $A_1$ when isovector correlator is considered \cite{ms-av}. 
Similarly from
\begin{eqnarray}
\frac{\Pi_1(q^2)-\Pi_1(0)}{q^2} = \frac{1}{\pi}\intop\frac{\Im\Pi_1(s) \ ds}{
s(s-q^2)} + {\rm subtractions}\nonumber\\ 
\label{dpi}
\end{eqnarray}
we obtain,
\begin{eqnarray}
\hat{L}_{M^2}\left[\frac{\Pi_1(q^2)}{q^2}\right]-\frac{\Pi_1(0)}{M^2} =
\frac{1}{\pi M^2}\intop \frac{\Im\Pi_1(s) \ e^{-s/M^2} \ ds}{s} \nonumber\\
\label{ldpi}
\end{eqnarray}
which can be used to calculate $\Pi_1(0)$. From the OPE expansion for the 
isovector current correlator Eqn.(38), we have
\begin{eqnarray}
\frac{\Pi_1(q^2)}{q^2} &=& -\frac{1}{4\pi^2} \ \left(1+\frac{\alpha_s}{\pi}
\right) \ ln \frac{Q^2}{\mu^2} - \frac{2}{Q^4} \ \hat{m}\langle\bar{q}q\rangle
\nonumber\\
&+&\frac{\alpha_s}{12\pi Q^4}\langle G^a_{\mu\nu}G^a_{\mu\nu}\rangle
+\left(\frac{32}{9}+\frac{64}{81}\right) \frac{\alpha_s\pi}{Q^6} \langle
\bar{q}q\rangle^2 \nonumber \\
\label{piope}
\end{eqnarray}
where $\hat{m}=(m_u+m_d)/2$. The above Eqn.(\ref{piope}) has also been used in 
the analysis of $\tau$ decay \cite{zya}. In the right hand side of
Eqn.(\ref{lpi}) and Eqn.(\ref{ldpi}) we use for the $A_1$, the central mass 
value from experiments and adopt two models
\begin{eqnarray}
{\rm model \ A} &:& \Im\Pi_1(s) = \pi h_A^2 m_{A_1}^4 \ \delta(s-m_{A_1}^2)\\
{\rm model \ B} &:& \Im\Pi_1(s) = \frac{K \ \Theta(s-(m_\rho+m_\pi)^2)}{
(s-m_{A_1}^2)^2 + \Gamma^2 m_{A_1}^2}
\end{eqnarray}
with $\Gamma = 300$ MeV. 
Following Zyablyuk \cite{zya} we use the values
\begin{equation}
g_s^2\langle G^a_{\mu\nu} G^a_{\mu\nu}\rangle = 0.5 \ ({\rm GeV})^4
\end{equation}
and for the four quark term
\begin{equation}
\frac{64\pi}{9} \alpha_s\langle \bar{q}q\rangle^2 = 3\times 10^{-3} \ 
({\rm GeV})^6
\label{q4}
\end{equation}
Eqn.(\ref{q4}) takes into account the renormalization corrections to the 4 quark
operator computed by Adam and Chetyrkin \cite{lea}. We note that the estimate in
eqn.(\ref{q4}) is smaller than the value used in the original work by Shiffman, 
Vainshtein and Zakharov \cite{ms-av}.
\begin{figure}
\begin{center}
\includegraphics[scale=0.60]{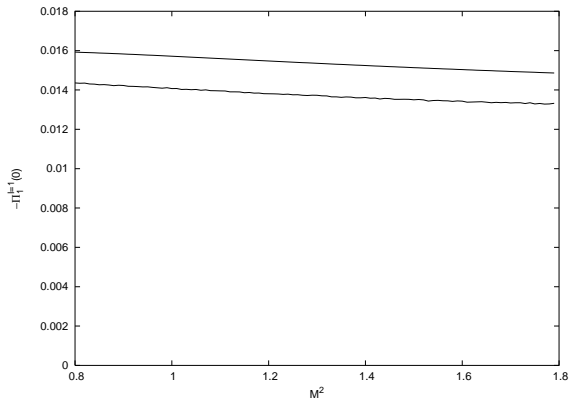}
\caption{\label{f1}
The plot show $-\Pi_1^{I=1}(0)$ obtained from Eqn.(48) in the interval $0.8 \le M^2
\le 1.8 (GeV)^2$. The upper curve is obtained for model A, cf. Eqn.(48), and 
the lower curve is for model B. See text.
}\end{center}
\end{figure} 

Using the above we find (cf. Fig.\ref{f1})
\begin{eqnarray}
\Pi_1(0) = -0.0154 \ (GeV)^2\hspace{2cm}{\rm (model \ A)}\\
\Pi_1(0) = -0.0138 \ (GeV)^2\hspace{2cm}{\rm (model \ B)}
\end{eqnarray}
which is to be compared with the ward identity value Eqn.(40) with $F_\pi^2 = 
0.017 \ (GeV)^2$. We like to stress that the above calculation does not use ward identity and PCAC.

Consider now the isoscalar case
\begin{eqnarray}
\Pi_{\mu\nu}^{I=0}&=&\frac{i}{2}\intop d^4x e^{i q.x} \langle0|T\{\bar{u} \gamma_\mu \gamma_5 u(x) + \bar{d}\gamma_\mu\gamma_5 d(x), \nonumber \\
&& \bar{u}\gamma_\nu\gamma_5 
u(0) + \bar{d}\gamma_\nu\gamma_5 d(0)\}|0\rangle
\end{eqnarray}
with
$$\Pi_{\mu\nu}^{I=0} = -\Pi^{I=0}_1(q^2) g_{\mu\nu}+\Pi^{I=0}_2(q^2)
q_\mu q_\nu.$$ 
Before proceeding with the calculations we note the interesting similarities
between the vector states $\rho(770), \ \omega(780), \ \phi(1020)$ and the axial
vector states $A_1(1235), f_1(1285), f_1(1420)$. Both $\rho$ and $A_1$ are broad
resonances and are nearly degenerate with their $I=0$ partners $\omega$ and
$f_1$ respectively. Morever the decays of  $\phi(1020), f_1(1420)$ are 
dominated by strange mesons. To evaluate $\Pi_1^{I=0}(0)$ we can proceed in an
analogous manner as for the isovector case above.
First we note that in the physical side or the r.h.s. of the sum
rules, Eqns.(46) and (48), $f_1(1285)$ will replace $A_1$. The higher mass state
$f_1(1420)$ is effectively included in the sum over higher mass states which in
the usual QCD sumrule approach are represented by the quark loop contributions
with an effective threshold $W^2$. We should stress here that the details of the
decay modes of $f_1(1285)$ and $f_1(1420)$ are irrelevant and do  not enter the
sum rules.

Now turning to the OPE it is clear, from Eqn.(38) and Eqn.(43), that as long 
as quark anihilation diagrams are neglected the OPE for $I=1$ and $I=0$ 
currents will be identical.  Consider the first term
in Eqn.(49). This arises from the single quark loop and is the same for the
isovector and isoscalar case. However in the isoscalar case we must include
corrections that can arise from a three loop diagram  in which the initial
quark current loop  annihilates to a two gluon intermediate followed by
materialisation to the final current quark loop.
These diagrams are expected to contribute with coefficients like 
$\alpha_s^2/\pi^2$ and for that reason we neglect them here.

\begin{figure}
\begin{center}
\includegraphics[scale=0.60]{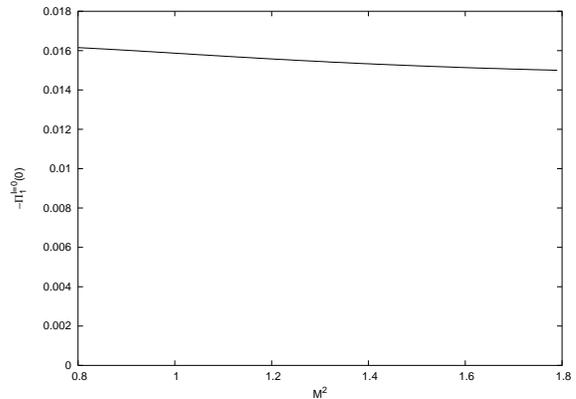}
\caption{\label{f2}
$\Pi_1^{I=0}(0)$ obtained from Eqn.(48) using $f_1(1285)$ and ignoring its width
as in model A. 
}\end{center}
\end{figure} 
The calculation of $\Pi_1^{I=0}(0)$ is then straight forward and we obtain (cf.
Fig.\ref{f2})
\begin{equation}
\Pi_1^{I=0}(0) = -0.0152 \ (GeV)^2
\end{equation}
In the analysis of the sumrule Eqn.(37) we shall therefore use $F = 0.0152 \ (GeV)^2$ as the central value and study the effect of variation around it.

We now turn to the determination of the dimension 5 correlator or the constant
$H$. For the isovector case this has been computed by Novikov et.al. 
\cite{van1}, using the sumrule for $\Pi_2$ in
\begin{eqnarray}
&i&\intop d^4x e^{iq.x} \langle0|T\{\bar{u}\gamma_\mu\gamma_5 d(x),g_s\bar{d}
\frac{\lambda^a}{2}\tilde{G}^a_{\beta\nu}\gamma_\beta u|0\rangle \nonumber\\
&=&-\Pi_1(q^2) g_{\mu\nu} + \Pi_2(q^2)q_\mu q_\nu
\end{eqnarray}
defining
\begin{equation}
\langle0|g_s \bar{d}\frac{\lambda^a}{2}\tilde{G}^a_{\beta\alpha}\gamma_\beta u(x)|\pi\rangle =
-\delta^2 F_\pi k_\alpha.
\end{equation}
They obtained $\delta^2 = 0.21 (GeV)^2$. However they had used a somewhat high 
value for the four quark correlator. This has been reanalysed by Ioffe and 
Oganesian \cite{ago} using more upto date values of various paramters. 
They obtain $\delta^2 = 0.16 (GeV)^2$. 
We  have independently reanalysed the sum rule of  Novikov et.al. \cite{van1} 
and agree with Ioffe and Oganesian \cite{ago}.
From Eqn(58) and Eqn(59) one gets the value $H(I=1)=\delta^2F_\pi^2$.

To find the value $H$ in the isoscalar case we proceed as follows. Ioffe
and Khodjamirian \cite{yu} have computed
\begin{equation}
\langle0|g_s\sum_q\bar{q}\frac{\lambda^a}{2}\tilde{G}^a_{\beta\alpha}
\gamma_\beta q(x)|0\rangle|_{SU(3) \ singlet} = 3 h_0 A_\alpha
\end{equation}
and find $h_0 \approx 3 \times 10^{-4}$ (GeV)$^4$.
Since the matrix elements in Eqns.(59) and (60) are quite small we can use 
SU(3) flavour symmetry and write
\begin{eqnarray}
&&\langle0|g_s\bar{q}\frac{\lambda^a}{2}\tilde{G}^a_{\nu\mu}\gamma_\nu 
q|0\rangle|_{Iso-singlet}\nonumber \\
&&=\frac{2}{3} \langle0|g_s\bar{q}\frac{\lambda^a}{2}
\tilde{G}^a_{\nu\mu}\gamma_\nu q|0\rangle|_{SU(3) \
singlet} \nonumber \\
&&+\frac{1}{3}\langle0|g_s\bar{q}\frac{\lambda^a}{2}\tilde{G}^a_{\nu\mu}
\gamma_\nu q|0\rangle |_{Octet}
\end{eqnarray}
From which we obtain 
\begin{eqnarray}
H &=& \frac{2}{3} h_0 + \frac{1}{3} \delta^2 F_{\pi}^2 \nonumber\\
&\approx& 1.14 \times 10^{-3} \ (GeV)^4
\end{eqnarray}
We shall use this as the central value in the our analysis of the sumrule
Eqn.(37). The above term is far less significant in the determinations of $G$
in Eqn.(\ref{rule}) than the dimension 3 
correlator constant $F$ calculated in the earlier part of this section.
\section{Determination of $G_U+G_D$ from the sumrule}
\label{five}
\begin{figure}
\begin{center}
\includegraphics[scale=0.60]{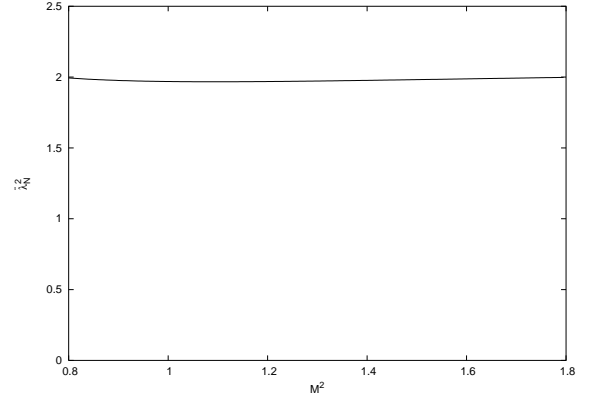}
\caption{\label{f3}
$\tilde{\lambda}^2_N$ obtained from Eqn.(63) is shown in the Borel mass interval 
$0.8 \le M^2 \le 1.8 \ (GeV)^2$.
}\end{center}
\end{figure}

\begin{figure}
\begin{center}
\includegraphics[scale=0.60]{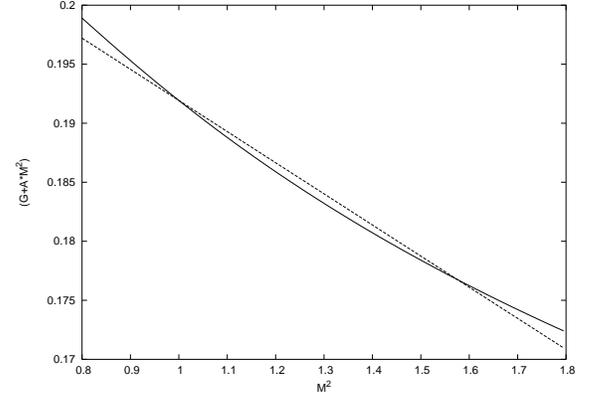}
\caption{\label{f4}
$G+AM^2$ obtained from Eqn.(37) is shown in the Borel mass interval $0.8 \le M^2
\le 1.8 \ (GeV)^2$. A straight line fit, dotted line gives $G=0.22$ and 
$A=-0.026$.
}\end{center}
\end{figure}

We can use the values of $F$ and $H$ determined in the previous section
in Eqn.(37) 
to find $G_U+G_D$. But first we need to determine $\tilde\lambda_N^2$ which can
be obtained from Ioffe's sumrule for the nucleon mass. 
\begin{eqnarray}
\frac{M^6 \ E_2}{L^{4/9}} + \frac{bM^2 \ E_0}{4L^{4/9}}
&+& \frac{4a^2L^{4/9}}{3} - \frac{a^2m_0^2}{3 M^2} \nonumber\\
&=& \tilde\lambda_N^2 \ e^{-m_N^2/M^2}
\end{eqnarray}  
Experimentally we have $\Lambda_{QCD}=247$ MeV [30],
which corresponds to $\alpha_s(1 \ {\rm GeV}) = 0.5$.
To find $\tilde\lambda_N^2$ we shall use the experimental value  $m_N=0.938$ GeV
and we fix $W^2 = 2.22$ by looking for the best fit in the least square sense in
the Borel mass interval 0.8 GeV$^2 < M^2 <$ 1.8 GeV$^2$, we find (cf.
Fig.\ref{f3})
$$ \tilde{\lambda}_N^2 = 1.975 \ GeV^6$$
This can now be used in the sumrule Eqn.(37), along with the values 
$F = 0.0152 \ (GeV)^2$ and  $H = 1.14\times10^{-3} \ (GeV)^4$. We find fitting 
the sum rule with $W^2 = 2.22$ and the same interval for $M^2$ as used in
Eqn.(63) (cf. Fig.\ref{f4})
\begin{equation}
 G = G_U+ G_D = 0.22 \ {\rm and} \ A = -0.026 \ (GeV)^{-2}
\end{equation}
Since our calculation of $F$ and $H$ in the previous section are not 
exact we have varied them by 20\%. An numerical increase of $F$ by 20\%
yields a value $G = 0.34$ and $A=-0.011 (GeV)^{-2}$, while a decrease by 20\%
gives a value $G= 0.10$ and $A=-0.041 (GeV)^{-2}$. Compared to this a change of
value of $H$ by 20\% barely changes the result by 2\%.
\section{conclusion}
\label{six}
To evaluate the linear combination $\frac{4}{9}G_U+\frac{1}{9}G_D + \frac{1}{9}
G_S$ occuring in the Bjorken sum rule we can proceed as follows. We write
\begin{eqnarray}
G_{Bj}&=& \frac{1}{6}(G_U-G_D) + \frac{1}{3}(G_U+G_D) \nonumber \\
&&- \frac{1}{18}(G_U+G_D-2G_S)
\end{eqnarray}
First term is known from neutron $\beta$ decay and we have
\begin{equation}
G_U-G_D = 1.267
\end{equation}
The last term in the Eqn.(59) is also known from hyperon $\beta$ decay assuming
the validity of flavour SU(3) symmetry. We have \cite{leader} 
\begin{equation}
G_U+G_D-2G_S = 0.585
\end{equation}
We can now use the value of $G_U+G_D$ determined in Sec.\ref{five}
Eqn.(64) to get 
\begin{equation}
G_{Bj}(\mu^2=1 \ GeV^2) = 0.32 \
\end{equation}
to be compared with experimental value \cite{adams}
\begin{equation}
G_{Bj}(\mu^2=5 \ GeV^2) \approx 0.28,
\end{equation}
which increases slightly when $\mu^2$ is decreased from 5 GeV$^2$ 
to 1 GeV$^2$. Taking into account the $\mu$ dependence of the singlet matrix
element \cite{t1,t3} the experimental number, Eqn.(69), increases by a few 
percent to 0.29 at $\mu^2 = 1 GeV^2$.

In arriving at Eqn.(68) we have used SU(3) flavour symmetry via Eqn.(67) in
Eqn.(65). Returning to Cabibbo theory, all the octet current matrix elements 
between the octet of baryon states are expressible in terms of two irreducible 
matrix elements F and D, and one has 
\begin{eqnarray}
      G_U - G_D   = F + D\\
      G_U + G_D - 2 G_S = 3F - D
\end{eqnarray}
In QCD sumrule calculations of the various octet current matrix elements 
between baryon states one first considers the
limit of massless quarks which is of course automatically SU(3) symmetric.
The constants $F$ and $D$ can be determined in a variety of ways. For example
the matrix elements of the isovector current $\bar{u}\gamma_\mu\gamma_5 u
-\bar{d}\gamma_\mu\gamma_5 d $ between nucleon states gives
$F + D$ while, $\Sigma\to\Sigma$ is proportional to $F$, $\Sigma\to\Lambda$ 
is proprtional to $D$ and $\Xi\to\Xi$ to $D - F$. Details can be found in 
ref.\cite{pisma,chiu}. In ref.\cite{chiu}, using a
different Lorentz structure invariant for the propagator in Eq.(26),
the relation  $7F\approx5D$ was found which is not far from experiment. To
incorporate the effect of quark masses one expands in quark mass and retains 
terms linear in them and also take into account the difference between
the strange quark condensate and up or down quark condensate. We first 
note that since up and down quark masses are negligible, the isovector current
matrix element in the nucleon is unaffected. In a recent update  of the earlier
calculation in ref.\cite{pisma}, Ioffe and Oganiesen \cite{ago} find
$$  G_U- G_D = 1.37 \pm 0.01$$
Finding $D$, $F$ and $D-F$ from the hyperon matrix elements, when quark masses
are included, is sensitively
dependent on the Borel mass region as was found by authors of \cite{chiu}. 
For this reason it is difficult to decide how accurately Cabbibo theory is
satisfied by using QCD sum rules. We have therefore relied on experiment which
is reflected in the use of Eqn.(67) to compute $G_{Bj}$ in Eqn.(65).
Nevertheless, it is worth emphasising that QCD sumrules provide, though not 
precise, a quantitative explanation of $G_{Bj}$, without any arbitrary 
parameter and use only the value of vacuum condensates already obtained 
through other hadron properties.
\appendix
\section{Wilson Coefficients}
There are two differences between the coefficients of the $M^2$ term in 
Eqn.(37),
namely, the coefficient of the gluon condensate term, $b$, and the coefficient
of the external field induced correlator, $H$, between ref.\cite{chiu,gupta,
henley} on the one hand and ref.\cite{bely,ago} on the other. Full
detail of the calculation can be found in the ref.\cite{chiu}. These 
calcualtions have since been verified again by the original authors of
ref.\cite{chiu} 
themselves and authors of ref.\cite{gupta} and ref.\cite{henley}. In
ref.\cite{chiu} the nucleon current
correlator, Eqn(26), is calculated for a generic $\Delta{\cal L}$
$$\Delta{\cal L} = (g_u \bar{u}\gamma_\mu \gamma_5 u + g_d \bar{d}\gamma_\mu
\gamma_5 d) A^\mu$$
with $g_u$ and $g_d$ as arbitray constants. According to Table 1 in ref. 
\cite{chiu}, the contribution of Fig (5) and (6) in ref.\cite{chiu} have
coefficient $-(2g_d + 10 g_u/3)$ and $2(g_u+g_d)$. In the computation of $G_A$,
$g_u = -g_d$ so that Fig (6) gives zero and there is complete agreement between
ref.\cite{chiu,henley} and ref.\cite{bely}. However for the isoscalar current, 
for
which $g_u = g_d = 1$, the results are different. Ref. \cite{chiu,henley} have
\begin{equation}
 -(2g_d + \frac{10}{3} g_u) + 2(g_u+g_d) = -\frac{4}{3}g_u 
\hspace{1cm}(g_u=g_d)
\label{a1}
\end{equation}
Reversing the sign of Fig (6) gives
\begin{equation}
 -(2g_d + \frac{10}{3} g_u)- 2(g_u+g_d) = -\frac{28}{3}g_u
\hspace{1cm}(g_u=g_d)
\end{equation}
which is result of ref.\cite{bely}. Thus ref.\cite{chiu} and \cite{bely} agree
numerically in the case $g_u =- g_d$ but different in the isoscalar case. In
Eqn.(37) the coeffcient used corresponds to Eqn.(\ref{a1}).

We now turn to the coefficient of the gluon condensate $\langle g_s^2
G_{\mu\nu}^a G_{\mu\nu}^a\rangle$. Again details of the calculations are given 
in ref.\cite{chiu} in their Table 1 which uses their eqn(2.12) and Fig.2 which 
leads to coefficient $-g_u/4$. On the other hand ref.\cite{bely} seem to have 
$g_d/4$ for this coefficient. Thus ref.\cite{chiu} and \cite{bely} agree
numerically in the case $g_u =- g_d$ but have opposite signs in the isoscalar
case. Interestingly enough if one tries to obtain the Wilson
coefficients for $T\{\eta(x)\bar{\eta}(0)\}|_{A_\nu}$ by using a chiral rotation
of the quark fields given by
\begin{eqnarray*}
u &\longrightarrow& e^{ig_u \ A.x \ \gamma_5} u\\
d &\longrightarrow& e^{ig_d \ A.x \ \gamma_5} d
\end{eqnarray*}
in the propagator $\langle0|T\{\eta(x)\bar{\eta}(0)\}|0\rangle$ in absence of
the external field, which occurs in the mass sum rule and assuming the validity
of chiral invariance of the vacuum to obtain
$\langle0|T\{\eta(x)\bar{\eta}(0)\}|_{A_\nu}|0\rangle$ one  
obtains $g_d/4$ for the coefficient of $\langle g_s^2G_{\mu\nu}^a 
G_{\mu\nu}^a\rangle$.
On the other hand it is well recognised that presence of non-perturbative gluon
fileds in the vacuum state leads to breakdown of chiral symmetry, which would
seem to invalidate such a calculation. Again in Eqn.(37) we have used the
result $-g_u/4$ found in ref.\cite{chiu} and ref.\cite{henley}.
Fortuitiously enough these two 
differences in the coefficients of third and fourth terms in Eqn.(37) between 
ref.\cite{chiu} and \cite{bely} numerically tend to compensate. Also we have 
seen the dimension three term characterised by $F$ which occures with $M^4$
in Eqn.(37), is far more significant than the $M^2$ term in obtaining $G_U+
G_D$. Our main point has been that $F$ can be computed from the spectrum of 
axial mesons and we get a sensible answer for $G_U+G_D$ and therefore $G_{Bj}$.

\begin{thebibliography}{30}
\bibitem{bj} J.D.Bjorken, Phys.Rev. {\bf 148}, 1467 (1966)
\bibitem{ej} J.Ellis and R.L.Jaffe, Phys.Rev. D {\bf 9}, 1444 (1974)
\bibitem{adams} D. Adams et.al. Phys. Rev. D {\bf 56} 5330 (1997)
\bibitem{leader} E. Leader and D. B. Stamenov, Phys.Rev. D {\bf 67}, 037503 
(2003)
\bibitem{gross} D.J.Gross, S.B.Treiman and F.Wilczek Phys.Rev. D {\bf 19} 2188 
1979
\bibitem{ioffe} B. L. Ioffe and A. V. Smilga Nucl. Phys. B {\bf 232}, 109 (1984)
\bibitem{iib} I. I. Balisky and A. V. Yung, Phys. Lett B {\bf129} 328 (1983)
\bibitem{pisma} V. M. Belyaev and Y. I. Kogan, Pism'a Zh. Eksp. Teor. Fiz.
{\bf 37} 611 (1983) [JETP Lett. {\bf37} 730 (1983)]; Phys. Lett. B {\bf136}, 
273 (1984)
\bibitem{chiu} C. B. Chiu, J. Pasupathy and  S. L. Wilson, Phys. Rev. D 
{\bf32}, 1786 (1985)
\bibitem{bely} V. M. Belyaev, B. L. Ioffe and Y. I. Kogan, Phys. Lett. B 
{\bf151}, 290 (1985)
\bibitem{gupta} S. Gupta, M. V. N. Murthy and J. Pasupathy, Phys. Rev. D 
{\bf39}, 2547 (1989)
\bibitem{henley} E. M. Henley, W-Y. P. Huang and L. S. Kisslinger, Phys. Rev.
D {\bf 46}, 431 (1992)
\bibitem{yu} B. L. Ioffe and A. Yu. Khodzhamirian, Sov. J. Nucl. Phys. 
{\bf55}, 1701 (1992)
\bibitem{bli} B. L. Ioffe, hep-ph/9804238
\bibitem{ago} B. L. Ioffe and A. G. Oganesian Phys. Rev. D {\bf57}, 6590 (1998)
\bibitem{van} V. A. Novikov et.al Nucl.Phys. B {\bf165}, 55 (1980)
\bibitem{shif} B. L. Ioffe and M. Shifman, Phys. Lett. B {\bf95}, 99 (1980)
\bibitem{gao} D. N. Gao, B. A. Li and M.-L. Yan, Phys. Rev. D {\bf56}, 4115 
(1997)
\bibitem{leut} H. Leutwyler, Phys. Lett. B {\bf378}, 313, (1996)
\bibitem{hrp} H. Pagels, Phys. Report C {\bf16}, 30 (1975)
\bibitem{sb} S. Brodsky, J. Ellis and M. Karlinev, Phys. Lett. B {\bf206}, 
309 (1988)
\bibitem{jp} J. Pasupathy, Phys. Rev. D {\bf12}, 2929 (1975), Phys. Lett. B 
{\bf48}, 71 (1975); C. A. Singh and J. Pasupathy, Phys.Rev. D {\bf18}, 
791 (1978)
\bibitem{see} See for example C. A. Dominguez and M. Loeve Phys.Rev. D {\bf31}, 
2930 (1985) and S. Narison, "QCD spectral sum rules", World scientific (1989)
\bibitem{ms-av} M. Shifman, A. Vainshtein and V. Zakharov, Nucl.Phy. B 
{\bf147}, 385,448 (1979); L. J. Reinders, H. Rubinstein and S. Yazaki, Phys.
Reports {\bf127}, 1 (1985)
\bibitem{zya} K. N. Zyablyuk, hpe-ph/0105346
\bibitem{lea} L. E. Adam and K. G. Chetyrkim Phys. Lett. B {\bf329}, 129 (1994)
\bibitem{van1} V.A.Novikov et.al Nucl.Phys B {\bf237}, 525 (1984)
\bibitem{t1} S. A. Larin, Phys. Lett. B {\bf303}, 113 (1993); K. G. Chetyrkin 
and J. H. K\"uhn, Z. Phys. C {\bf60}, 497 (1993)
\bibitem{t3} A. L. Kateav, Phys. Rev. D {\bf50}, R5469 (1994)
\end{thebibliography}
\end{document}